\documentclass[12pt,english]{paper}
\usepackage{amssymb}
\usepackage[T1]{fontenc}
\usepackage[latin1]{inputenc}
\usepackage{graphicx}
\usepackage{cite}
\makeatletter

\makeatletter

\makeatletter

\usepackage{geometry}

\makeatletter

\usepackage{epsfig}

\makeatother

\makeatother

\makeatother

\usepackage{babel}
\makeatother
\begin{document}

\title{Directly search for Ultra-High Energy WIMPs at IceCube}

\author{Ye Xu$^{1,2}$}

\maketitle

\begin{flushleft}
$^1$School of Information Science and Engineering,  Fujian University of Technology, Fuzhou 350118, China
\par
$^2$Research center for Microelectronics Technology, Fujian University of Technology, Fuzhou 350118, China
\par
e-mail address: xuy@fjut.edu.cn
\end{flushleft}

\begin{abstract}
I study the possibility of directly detecting Ultra-high energy (UHE from now on) WIMPs by the IceCube experiment, via the WIMPs interaction with the nuclei in the ice. I evaluate galactic and extragalactic UHE WIMP and astrophysical and atmospheric neutrino event rates at energy range of 10 TeV - 10 PeV. I assume UHE WIMPs $\chi$ are only from the decay of superheavy dark matter $\phi$, that is $\phi\to\chi\bar{\chi}$. If the lifetime of superheavy dark matter $\tau_{\phi}$ is taken to be $5\times10^{21}$s, WIMPs can be detected at energies above O(40-100)TeV in this detection. Since UHE WIMP fluxes are actually depended on $\tau_{\phi}$, if superheavy dark matter can decay to standard model particles (that is $\tau_{\phi}$ is constrained to be larger than O($10^{26}$-$10^{29}$)s), UHE WIMPs could not be detected by IceCube.
\end{abstract}

\begin{keywords}
UHE WIMPs, Superheavy dark matter, Direct detection of dark matter
\end{keywords}

\section{Introduction}
It is indicated by the Planck data with measurements of the cosmic microwave background
 that $26.6\%$ of the overall energy density of
the Universe is non-baryonic dark matter \cite{Planck2015}. Weakly Interacting Massive Particles (WIMPs from now on),
predicted by extensions of the Standard Model of particle physics,
are a class of candidates for dark matter\cite{GDJ}. They are
distributed in a halo surrounding a galaxy. At present, one mainly searches for thermal
WIMPs via direct and indirect detections\cite{CDMSII,CDEX,XENON1T,LUX,PANDAX,AMS-02,DAMPE,fermi}. Because of the very small cross sections of the interactions
between the thermal WIMPs and nucleus (maybe O(10$^{-47}$ cm$^2$))\cite{XENON1T,PANDAX}, so far one has not found dark matter yet.
\par
It is a reasonable assumption that there exist various dark matter particles in the Universe. Then it is possible that
this sector may comprise of non-thermal components. And these particles may also contain a small
component which is relativistic. Although the fraction of these relativistic dark matter particles
is small in the Universe, their large interaction cross sections (including between themselves and between them and the
Standard Model (SM from now on) particles) make it possible to search for them. Due to the reasons mentioned above, one has
to shift more attention to direct and indirect detection of UHE WIMPs. In fact, one has discussed the possibility of searching for UHE dark matter particles in Ref.\cite{ACNT,BGA,xu1}.
\par
There is a non-thermal dark sector generated by the early Universe with its bulk comprised of a very massive
relic $\phi$ in the Universe. This superheavy dark matter\cite{KC87, CKR98, CKR99, CGIT, FKMY} decays to another much lighter WIMPs $\chi$ and its lifetime is greater
than the age of the Universe. This lead to a small but significant flux of relativistic WIMPs\cite{LT, BGA, EIP, BLS, BKMTZ}. Actually, a few anomalous events may be UHE signatures of a beyond SM particle in the recent observations from the IceCube and ANITA experiments\cite{anita,icecube2018}.
\par
In the present work, I only focus on direct detection of the UHE WIMPs $\chi$ induced by the decay of superheavy dark matter $\phi$ ($\phi\to\chi\bar{\chi}$). These relativistic WIMPs $\chi$, which pass through the Earth and ice and interact with nuclei, can be detected by a detector like IceCube. In this detection, the main contamination is from astrophysical and atmospheric neutrinos.
\par
In what follows, I will estimate the UHE WIMP and neutrino event rates at IceCube. And it is discussed the possibility of direct detection of UHE WIMPs induced by the decay of superheavy dark matter at IceCube.
\section{UHE WIMPs flux from the Galaxy and Extra Galaxy}
I consider a scenario where the dark matter sector is composed of at least two particle species in the Universe. One is a co-moving non-relativistic scalar species $\phi$, with mass $m_{\phi} >$ 10 TeV, the other is much lighter particle species $\chi$ ($m_{\chi} \ll m_{\phi}$), due to the decay of $\phi$, with a very large lifetime. And $\chi$ comprises the bulk of present-day dark matter. Since I assume the superheavy dark matter $\phi$ does not decay to SM particles, the constraint on the $\phi$ lifetime are only those based on cosmology\cite{IOT,NS,DMQ,popolo}, that is $\tau_{\phi}\geq10^{17}$s. In the present work, $\tau_{\phi}$ is taken to be $5\times10^{21}$s\cite{BGA}.
\par
The UHE WIMPs flux is composed  of galactic and extragalactic components. So the total flux $\psi_{\chi}=\psi_{\chi}^G+\psi_{\chi}^{EG}$, where $\psi_{\chi}^G$ and $\psi_{\chi}^{EG}$ are the galactic and extragalactic fluxes, respectively.  The UHE WIMPs flux from the Galaxy is obtained via the following equation\cite{BGA,EIP}:
\begin{center}
\begin{equation}
\psi_{\chi}^G=F^G\int_{E_{min}}^{E_{max}}\frac{dN_\chi}{dE_\chi}dE
\end{equation}
\end{center}
with
\par
\begin{center}
\begin{equation}
F^G=1.7\times10^{-8}cm^{-2}s^{-1}\times\frac{10^{26}s}{\tau_{\phi}}\times\frac{1TeV}{m_{\phi}} cm^{-2}s^{-1}sr^{-1}.
\end{equation}
\end{center}
Its zenith angle distribution is similar to the one of the galactic UHE neutrino flux at IceCube(see Fig.1 in Ref.\cite{CIMM}) in the present paper.
\par
The UHE WIMPs flux from the extra galaxy is obtained via the following equation\cite{BGA,EIP}:
\begin{center}
\begin{equation}
\psi_{\chi}^{EG}=F^{EG}\int_{E_{min}}^{E_{max}}dE \int_0^{\infty}dz\frac{1}{\sqrt{\Omega_{\Lambda}+\Omega_m(1+z)^3}}\frac{dN_\chi}{dE_\chi}[(1+z)E_\chi]
\end{equation}
\end{center}
with
\par
\begin{center}
\begin{equation}
F^{EG}=1.4\times10^{-8}cm^{-2}s^{-1}\times\frac{10^{26}s}{\tau_{\phi}}\times\frac{1TeV}{m_{\phi}} cm^{-2}s^{-1}sr^{-1}.
\end{equation}
\end{center}
where z represents the red-shift of the source, $\Omega_{\Lambda}=0.685$ and $\Omega_m=0.315$ from the PLANCK experiment\cite{Planck2015}. $\displaystyle\frac{dN_{\chi}}{dE_{\chi}}=\delta(E_{\chi}-\displaystyle\frac{m_{\phi}}{2})$, where E$_{\chi}$ and N$_{\chi}$ are the energy and number of UHE WIMP, respectively.
\section{UHE WIMP and neutrino interactions with nuclei}
In the present paper, I take a Z' portal dark matter model\cite{APQ,Hooper} for WIMPs to interact with nuclei within the IceCube zone (see Fig.1). In this model, a new Z' gauge boson is considered as a simple and well-motivated extension of SM (see Fig.1(a) in Ref.\cite{BGA}). And the parameters in the model are taken to be the same as the ones in Ref.\cite{BGA}, that is, the interaction vertexes ($\chi\chi$Z' and qqZ') are assumed to be vector-like, the coupling constant G ($G=g_{\chi\chi Z}g_{qqZ}$) is chosen to be 0.05 and the Z' and $\chi$ masses are taken to be 5 TeV, 10 GeV, respectively. Theoretical models that encompass WIMP spectrum have been discussed in the literature in terms of Z or Z' portal sectors with Z' vector boson typically acquiring mass through the breaking of an additional U(1) gauge group at the high energies (see Ref.\cite{APQ,Hooper}). The UHE WIMP interaction cross section with nucleus is obtained by the following function(see Fig.1(b) in Ref.\cite{BGA}):
\begin{center}
\begin{equation}
\sigma_{\chi N}=6.13\times10^{-43} cm^2 \left(\frac{E_{\chi}}{1GeV}\right)^{0.518}
\end{equation}
\end{center}
\par
For neutrinos at the energy range of 10 TeV - 10 PeV, the interaction cross sections with nucleus are are given by simple power-law forms from Ref.\cite{BHM}:
\begin{center}
\begin{equation}
\sigma_{\nu N}(CC)=1.17\times10^{-36} cm^2 \left(\frac{E_{\nu}}{1 GeV}\right)^{0.459}
\end{equation}
\end{center}
\begin{center}
\begin{equation}
\sigma_{\nu N}(NC)=3.78\times10^{-37} cm^2 \left(\frac{E_{\nu}}{1 GeV}\right)^{0.472}
\end{equation}
\end{center}
where $E_{\nu}$ is the neutrino energy.
\par
Then the above equations show that $\sigma_{\chi N}$ is smaller by 6-7 orders of magnitude, compared to $\sigma_{\nu N}$, at the same energies. The WIMP and neutrino interaction length can be obtained by
\par
\begin{center}
\begin{equation}
L_{\nu,\chi}=\frac{1}{N_A\rho\sigma_{\nu,\chi N}}
\end{equation}
\end{center}
\par
where $N_A$ is the Avogadro constant, and $\rho$ is the density of matter, which WIMPs and neutrinos interact with.
\section{Evaluation of the numbers of UHE WIMPs and neutrinos detected by IceCube}
IceCube is a km$^3$ neutrino telescope and can detect three flavour neutrinos via detecting the secondary particles, that in turn emit Cherenkov photons, produced by the interaction between neutrinos and the Antarctic ice\cite{icecube2018}. UHE WIMPs reach the Earth and pass through the Earth and ice, meanwhile these particles interact with matter of the Earth and ice. Cherenkov Photons are produced by cascades due to UHE WIMPs interaction with nuclei within the IceCube (see Fig. 1). A small part of these photons will be detected by the IceCube detector. Since the UHE WIMP interact with the nucleus in the ice and this is very similar to deep inelastic scattering, its secondary particles develop into a cascade at IceCube. The numbers of UHE WIMPs and neutrinos, (N$_{det}$)$_{\chi,\nu}$, detected by IceCube can be obtained by the following function:
\begin{center}
\begin{equation}
(N_{det})_{\chi,\nu} = R\times T\times \int_{E_{min}}^{E_{max}} \int_{cos\theta_{max}}^1(A\Omega)_{eff}(cos\theta) \Phi_{\chi,\nu}(cos\theta,E)dcos\theta dE
\end{equation}
\end{center}
where R and T are the duty cycle and lifetime of the IceCube experiment, respectively. $\theta$ is the polar angle for the Earth (see Fig. 1),  and $\theta_{max}$ is the maximum of $\theta$. $\Phi_\chi=\displaystyle\frac{d\psi_\chi}{dE}$ and $(A\Omega)_{eff}(cos\theta)$ = the observational area $\times$ the effective solid angle $\times$ P(E,$D_e(cos\theta)$,D). Here
\begin{center}
\begin{equation}
P(E,D_e(cos\theta),D)=exp\left(-\displaystyle\frac{D_e(cos\theta)}{L_{earth}}\right)\left[1-exp\left(-\displaystyle\frac{D}{L_{ice}}\right)\right]
\end{equation}
\end{center}
where $P(E,D_e(cos\theta),D)$ is the probabilities that UHE WIMPs and neutrinos interact with ice after traveling a distance between $D_e(cos\theta)$ and $D_e(cos\theta)+D$, where D is the effective length in the IceCube detecting zone in the ice, $D_e(cos\theta)$ are the distances through the Earth, and $L_{earth,ice}$ are the UHE WIMP and neutrino interaction lengths with the Earth and ice, respectively. In the present paper, I assume that all UHE WIMPs and neutrinos detected by IceCube interact with the ice within its volume.
In what follows, I will make a rough study of $(A\Omega)_{eff}(cos\theta)$ and then evaluate the numbers of UHE WIMPs detected by IceCube. Here I make an assumption that the observational area of IceCube is regarded as a point in the calculation of the effective solid angle $\Omega$. Under this approximation,
\begin{center}
\begin{equation}
(A\Omega)_{eff}(cos\theta) \approx P(E,D_e(cos\theta),D)A\frac{2\pi {R_e}^2}{D_e(cos\theta)^2} .
\end{equation}
\end{center}
where A is the observational area, $R_e$ is the radius of the Earth,  $D_e(cos\theta)=\displaystyle\frac{R_e(1+cos\theta)}{cos\theta_z}$. $\theta_z$ is the zenith angle at IceCube and $\theta_z=\theta/2$. And the observational area $\approx 1 km^2$, D $\approx$ 1 km and R $\approx$ 100\% at IceCube.
\par
The background is mainly two sources: astrophysical and atmospheric neutrinos. The astrophysical neutrinos is estimated with a diffuse neutrino flux of $\Phi_{\nu}=0.9^{+0.30}_{-0.27}\times(E_{\nu}/100TeV)\times10^{-18}GeV^{-1} cm^{-2}s^{-1}sr^{-1}$\cite{icecube}, where $\Phi_{\nu}$ represents the per-flavor flux, by the above method. And The atmospheric neutrinos is estimated with a flux of $lg E_{\nu}^2\Phi_{\nu}=\sum\limits_{k=0}^4\sum\limits_{n=0}^3a_{kn}x^ny^k$ by the same method\cite{SPS}, where $\Phi_{\nu}$ represents the atmospheric neutrino flux, $y=lg(E_{\nu}/1GeV)$, $x=cos\theta_z$ and coefficients $a_{kn}$ are given in Table 1 in Ref.\cite{SPS}.
\par
There is the presence of the Glashow resonance(on-shell $W^-$ via $\bar{\nu}_e+e^- \to W^- \to anything$ and its cross section is $5.02\times10^{-31}cm^2$) at 6.3 PeV and its secondary particles develop into cascades at IceCube\cite{SLG,BG1,BG2,BGRW}. So this effect has to be taken into account in the background estimation.
\par
A mixture of all flavours of UHE neutrinos consists of two different event topologies(cascades and tracks) at IceCube. We can reject a part of neutrinos by distinguishing track-like events from cascade-like events. So the neutrino contamination is actually less than the results of the calculation in the present paper. As the conservative estimation of background, however, I didn't consider this event topology identification in my calculation yet.
\section{Results}
In my work, the UHE WIMP and neutrino event rates are evaluated at energy range of 10 TeV - 10 PeV at IceCube. Fig. 2 - 7 compare the galactic and extragalactic UHE WIMP event rates to the astrophysical and atmospheric neutrino event rates corresponding to different $\theta_{max}$ when $\tau_{\phi}=5\times10^{21}$s, respectively. We find that an energy threshold has to be set to reject neutrinos, since neutrinos are more dominant in the events detected by IceCube at low energy(see Fig. 2 - 7).
\par
Fig. 2 shows the WIMP and neutrino event rates at $\theta_{max}=\displaystyle\frac{\pi}{6}$. In Fig. 2, we find that UHE WIMPs can be detected by IceCube at energies above about 40 TeV, and their event rates are $\sim$2 event/year and $\sim$2 events/10 years at 40 TeV and 10 PeV, respectively. Fig. 3 shows the WIMP and neutrino event rates at $\theta_{max}=\displaystyle\frac{\pi}{3}$. In Fig. 3, we find that UHE WIMPs can be detected by IceCube at energies above about 100 TeV, and their event rates are $\sim$10 events/year and $\sim$1 event/year at 100 TeV and 10 PeV, respectively. Fig. 4 shows the WIMP and neutrino event rates at $\theta_{max}=\displaystyle\frac{\pi}{2}$. In Fig. 4, we find that UHE WIMPs can be detected by IceCube at energies above about 110 TeV, and their event rates are $\sim$24 events/year and $\sim$3 events/year at 110 TeV and 10 PeV, respectively. Fig. 5 shows the WIMP and neutrino event rates at $\theta_{max}=\displaystyle\frac{2\pi}{3}$. In Fig. 5, we find that UHE WIMPs can be detected by IceCube at energies above about 130 PeV, and their event rates are $\sim$40 events/year and $\sim$5 events/year at 130 TeV and 10 PeV, respectively. Fig. 6 shows the WIMP and neutrino event rates at $\theta_{max}=\displaystyle\frac{5\pi}{6}$. In Fig. 6, we find that UHE WIMPs can be detected by IceCube at energies above about 160 TeV, and their event rates are $\sim$63 events/year and $\sim$9 events/year at 160 TeV and 10 PeV, respectively. Fig. 7 shows the WIMP and neutrino event rates at $\theta_{max}=\pi$. In Fig. 7, we find that UHE WIMPs can be detected by IceCube at energies above about 380 TeV, and their event rates are $\sim$925 events/year and $\sim$204 events/year at 380 TeV and 10 PeV, respectively.
\section{Conclusion and Discussion}
According to the results described above, it is possible that UHE WIMPs are directly detected with a detector like IceCube. Especially, the WIMPs with O(40-100)TeV from the other side of the earth (that is $\theta_{max} < \displaystyle\frac{\pi}{2}$) can be detected by IceCube when $\tau_{\phi}=5\times10^{21}$s.
\par
Since the UHE WIMP flux $\Phi_{\chi}$ is proportional to $\displaystyle\frac{1}{\tau_{\phi}}$, WIMP event rates are actually depended on the lifetime of superheavy dark matter. If we assume the superheavy dark matter $\phi$ can decay to SM particles, $\tau_{\phi}$ is constrained to be larger than O($10^{26}$-$10^{29}$)s\cite{EIP,MB,RKP,KKK} by diffuse gamma-ray and neutrino observations. Fig. 8-10 show the WIMP and neutrino event rates are evaluated at $\theta_{max}=\pi$ when $\tau_{\phi}=10^{17}$, $10^{25}$ and $10^{26}$s, respectively. In Fig. 8, we find UHE WIMPs can be detected at energies 10 TeV - 10 PeV at IceCube when $\tau_{\phi}=10^{17}$s. In Fig. 10, We find that WIMPs could not be detected at energies 10 TeV - 10 PeV since neutrinos are more dominant in the events by IceCube when $\tau_{\phi}$ is larger than $10^{26}$s.
\par
In Fig. 9, we find the WIMPs can be detected by IceCube only at O(1PeV) and their event rate is O(1) events/10 years when $\tau_{\phi}=10^{25}$s after deducting the Glashow resonance contamination at near 6.3 PeV. This is consistent with the observations that there are three anomalous events at O(1PeV) in nine years of IceCube data\cite{icecube2018,anita}. If they are actually UHE WIMPs, the lifetime of superheavy dark matter may be $\sim$10$^{25}$s. Fig. 11 shows the Signal to Background Rates(SBR) at different energy and different $\theta$ when $\tau_{\phi}=10^{25}$s. $SBR=\displaystyle\frac{N_{WIMP}}{N_{\nu}}$, where $N_{WIMP}$ and $N_{\nu}$ are the event rates of UHE WIMPs and neutrinos detected at IceCube, respectively. In Fig. 11, we find UHE WIMPs can be detected at energies above $\sim$2 PeV and the neutrino contamination is down to O($10^{-5}$) at low $\theta$ and high energy region(about $0^{\circ} < \theta < 65^{\circ}$ and 8 PeV < E < 10 PeV). It is significantly for the Glashow resonance to make an effect on this WIMP detection at near 6.3 PeV. There is the peak due to the UHE WIMP flux of the galactic center at about 56$^\circ$ in Fig. 11.
\section{Acknowledgements}
This work was supported by the National Natural Science Foundation
of China (NSFC) under the contract No. 11235006, the Science Fund of
Fujian University of Technology under the contract No. GY-Z14061 and the Natural Science Foundation of
Fujian Province in China under the contract No. 2015J01577.
\par

\newpage

\begin{figure}
 \centering
 \includegraphics[width=0.9\textwidth]{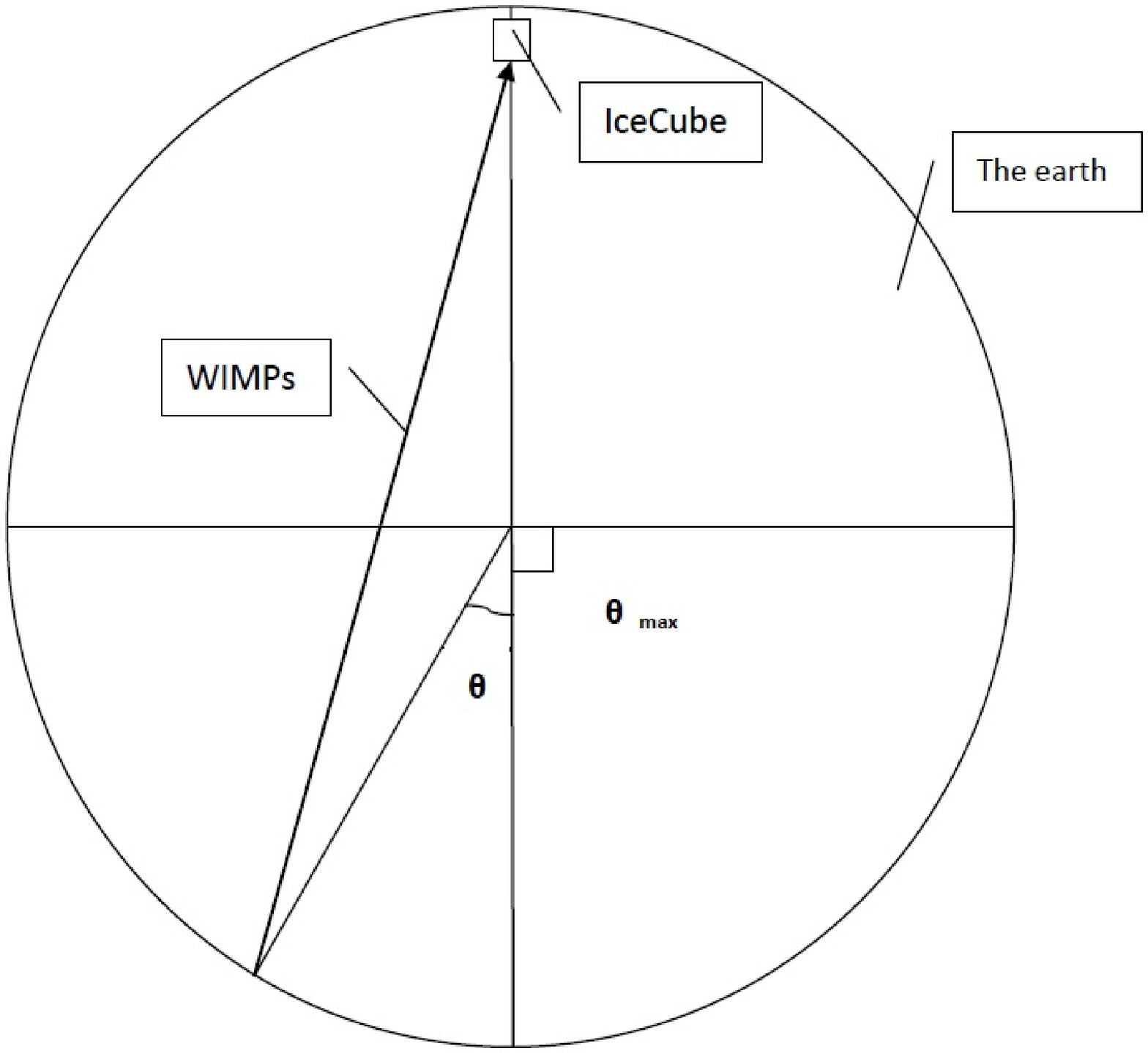}
 \caption{UHE WIMPs pass through the Earth and ice and can be detected by a detector like IceCube, via Cherenkov photons due to the development of cascades in the ice. $\theta$ is the polar angle for the Earth.$\theta_{max}$ is the maximum of $\theta$}
 \label{fig:figure}
\end{figure}

\begin{figure}
 \centering
 \includegraphics[width=0.9\textwidth]{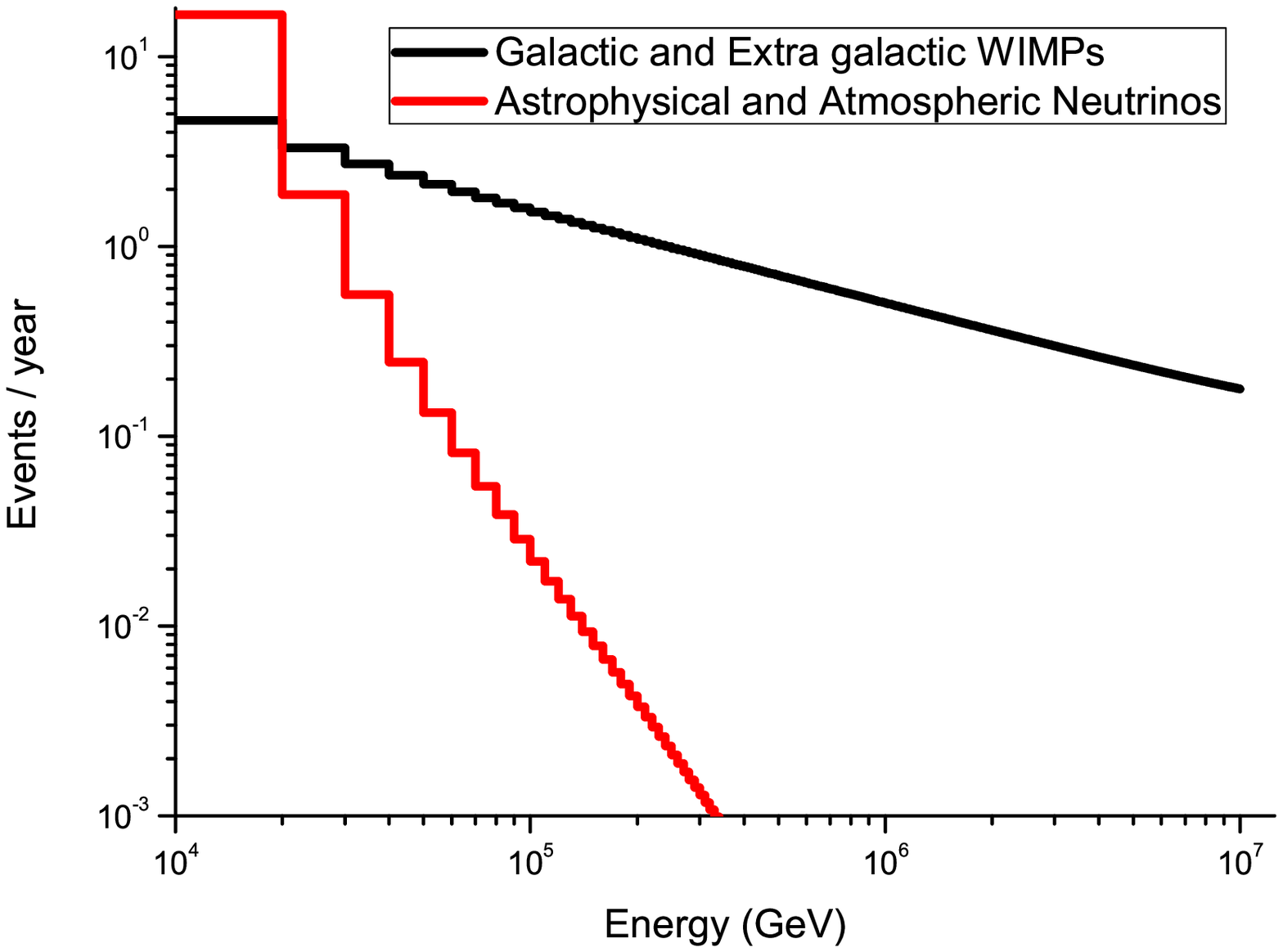}
 \caption{The UHE WIMPs and neutrino event rates are evaluated at $\theta_{max} = \displaystyle\frac{\pi}{6}$ at IceCube}
 \label{fig:30}
\end{figure}

\begin{figure}
 \centering
 \includegraphics[width=0.9\textwidth]{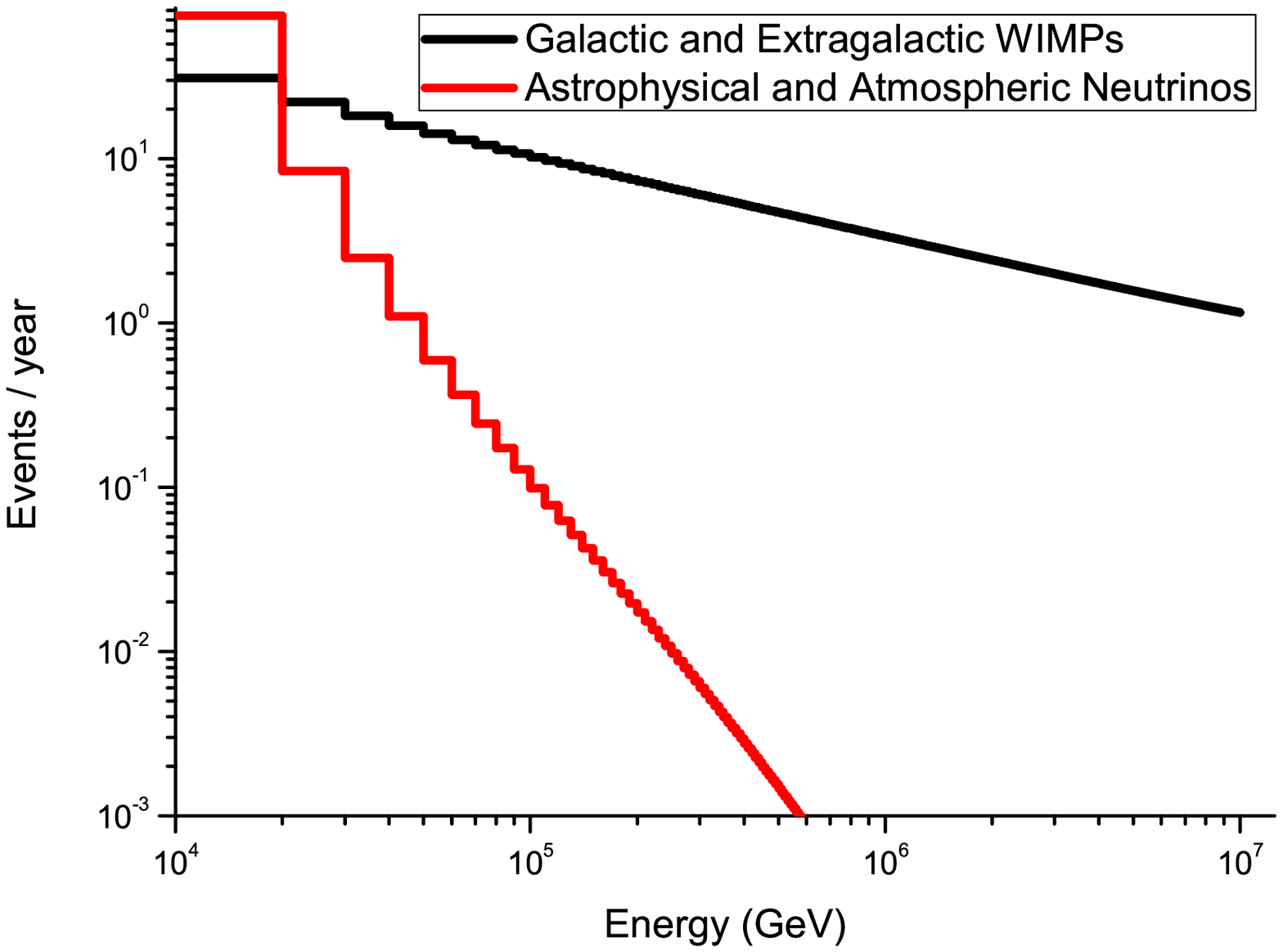}
 \caption{The UHE WIMPs and neutrino event rates are evaluated at $\theta_{max} = \displaystyle\frac{\pi}{3}$ at IceCube}
 \label{fig:60}
\end{figure}

\begin{figure}
 \centering
 \includegraphics[width=0.9\textwidth]{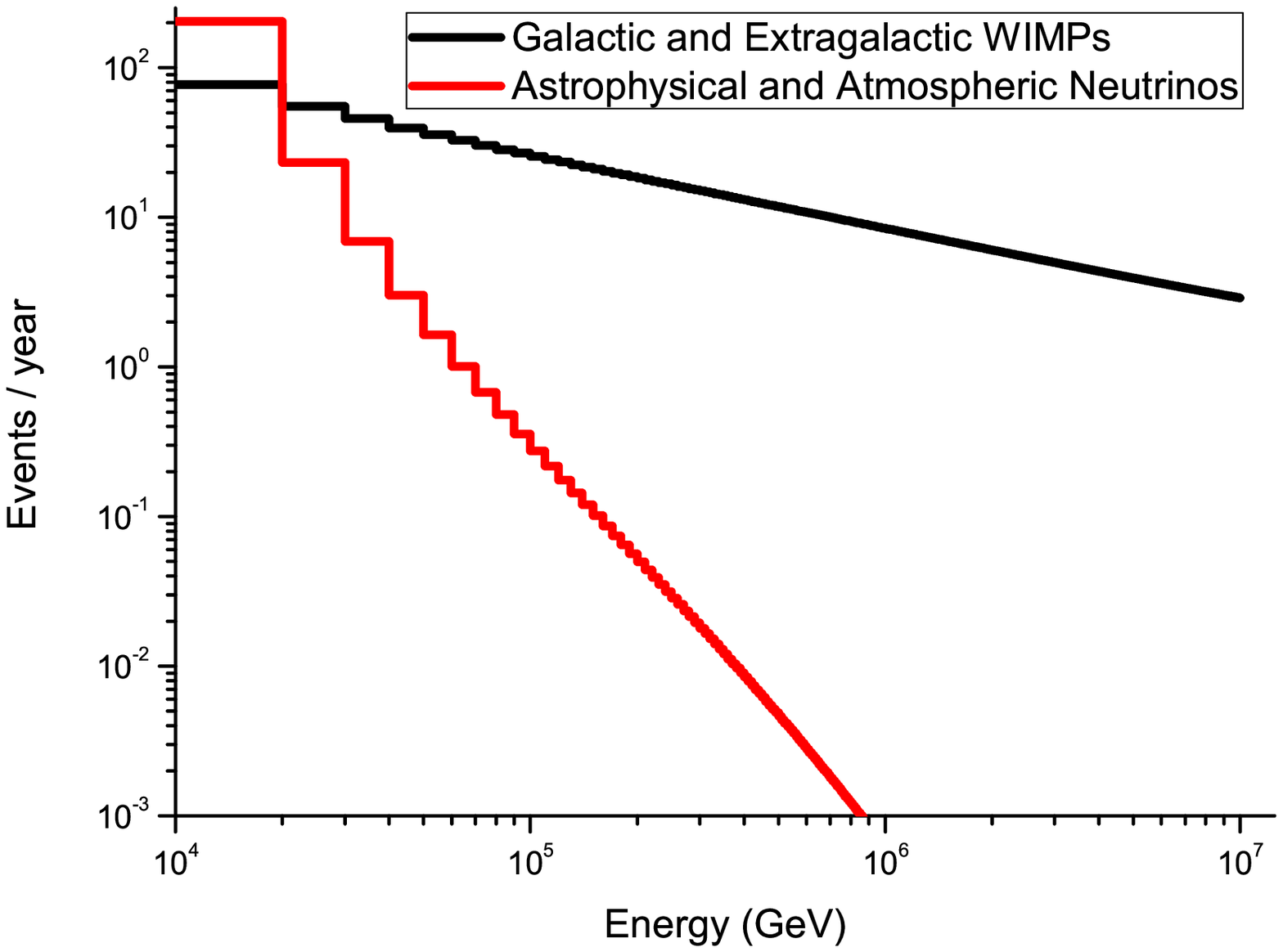}
 \caption{The UHE WIMPs and neutrino event rates are evaluated at $\theta_{max} = \displaystyle\frac{\pi}{2}$ at IceCube}
 \label{fig:90}
\end{figure}

\begin{figure}
 \centering
 \includegraphics[width=0.9\textwidth]{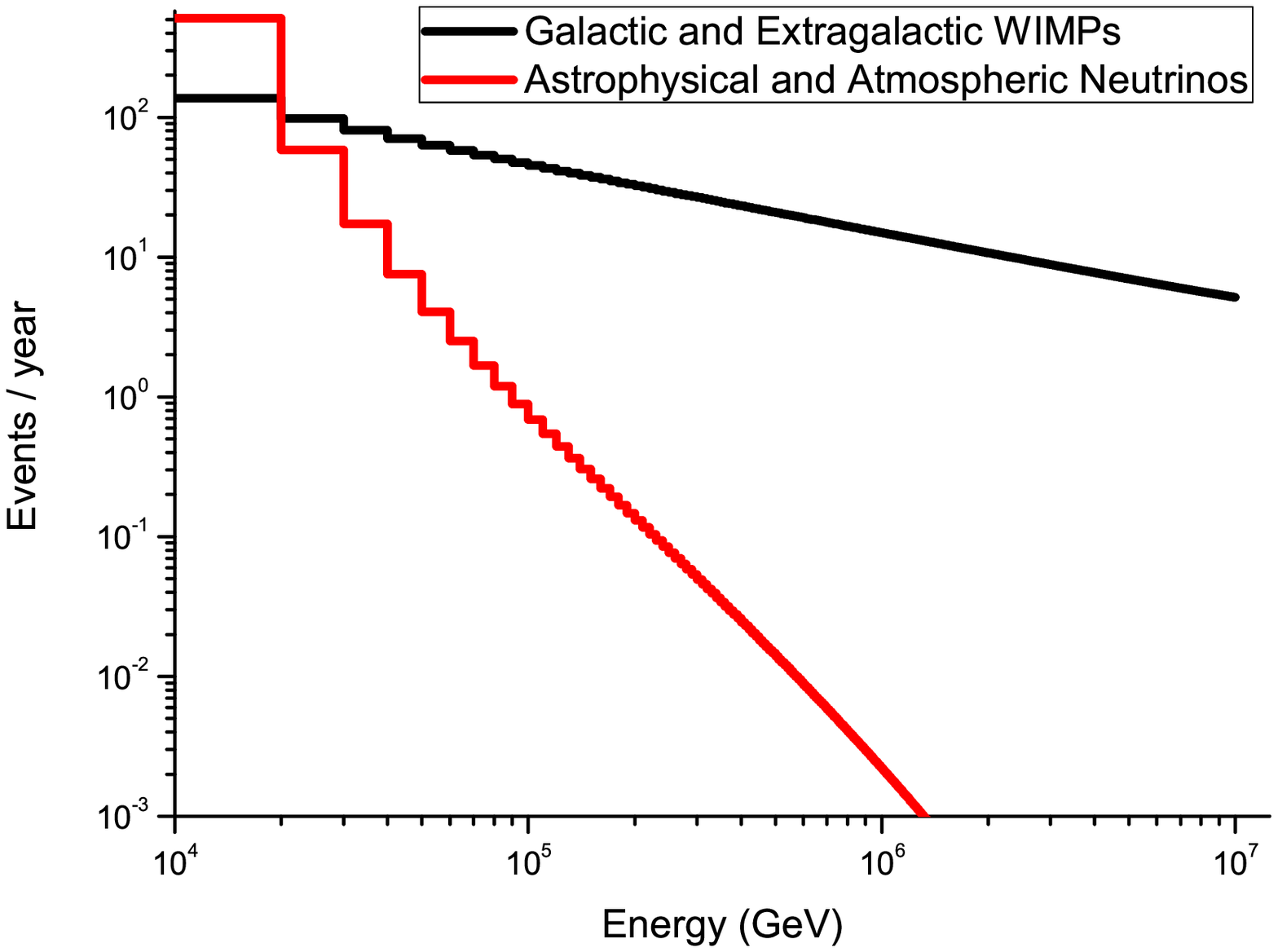}
 \caption{The UHE WIMPs and neutrino event rates are evaluated at $\theta_{max} = \displaystyle\frac{2\pi}{3}$ at IceCube}
 \label{fig:120}
\end{figure}

\begin{figure}
 \centering
 \includegraphics[width=0.9\textwidth]{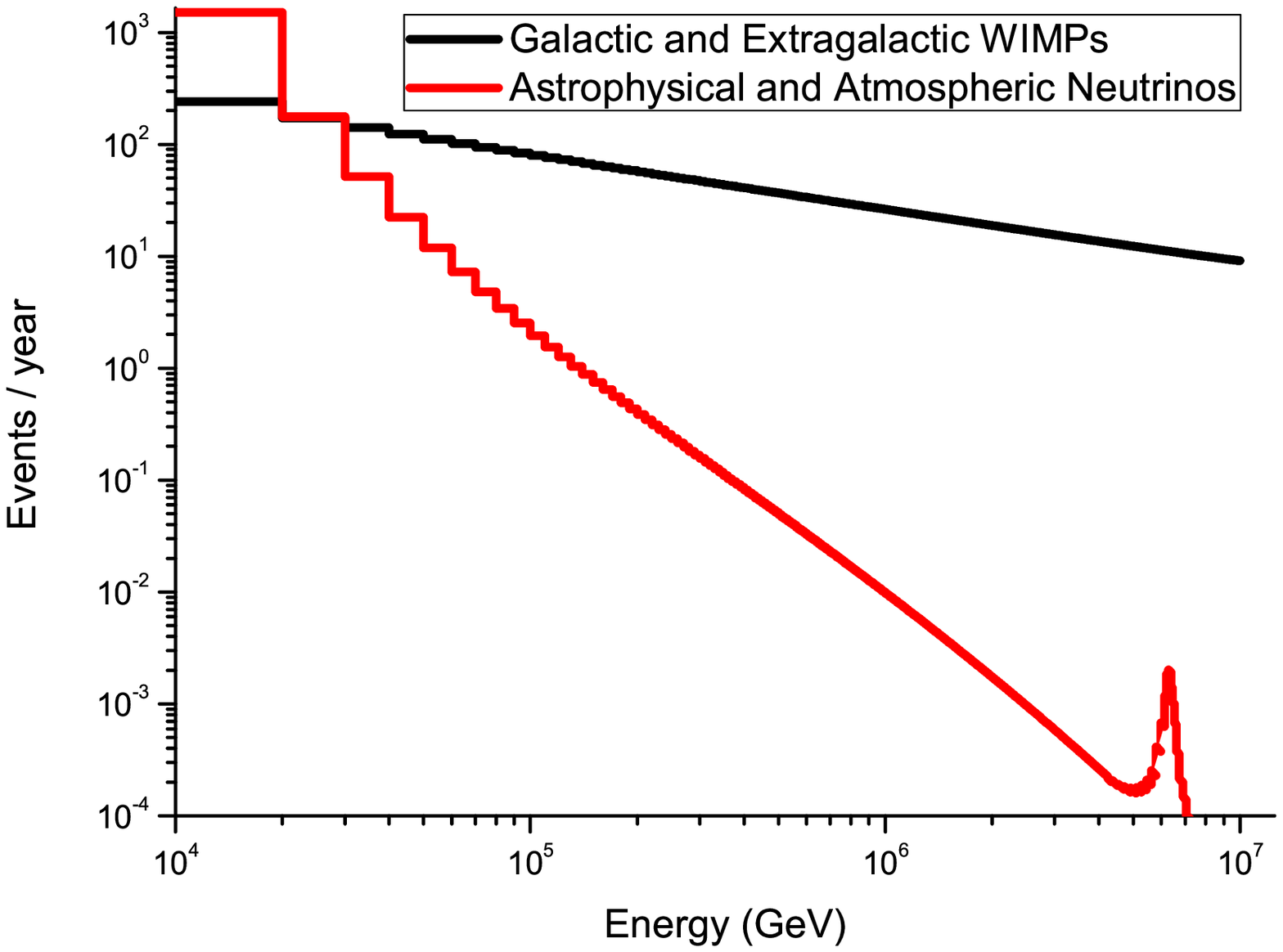}
 \caption{The UHE WIMPs and neutrino event rates are evaluated at $\theta_{max} = \displaystyle\frac{5\pi}{6}$ at IceCube}
 \label{fig:150}
\end{figure}

\begin{figure}
 \centering
 \includegraphics[width=0.9\textwidth]{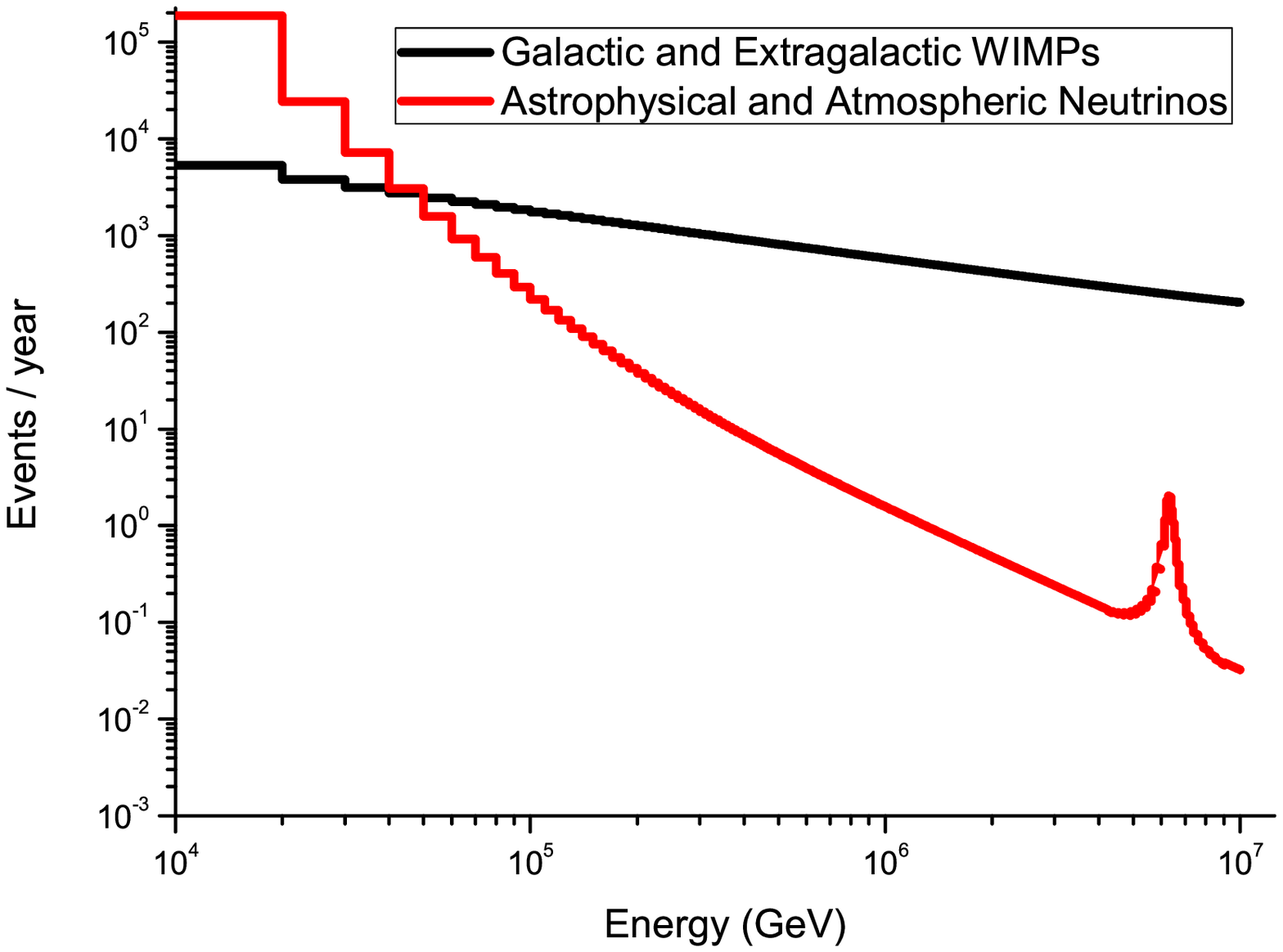}
 \caption{The UHE WIMPs and neutrino event rates are evaluated at $\theta_{max} = \pi$ at IceCube}
 \label{fig:180}
\end{figure}

\begin{figure}
 \centering
 \includegraphics[width=0.9\textwidth]{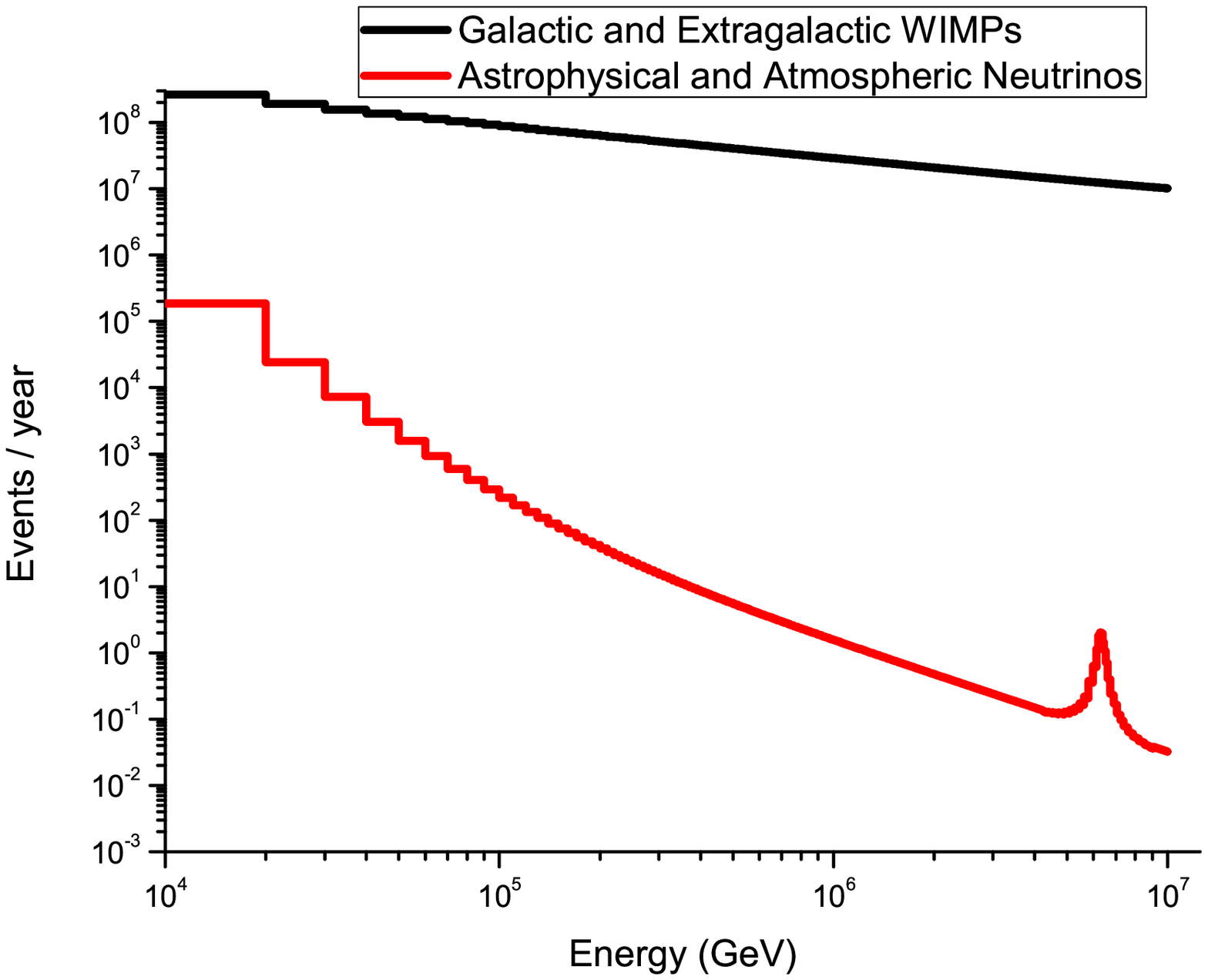}
 \caption{The UHE WIMPs and neutrino event rates are evaluated at $\theta_{max} = \pi$ at IceCube, if $\tau_{\phi}=10^{17}s$.}
 \label{fig:180_17}
\end{figure}

\begin{figure}
 \centering
 \includegraphics[width=0.9\textwidth]{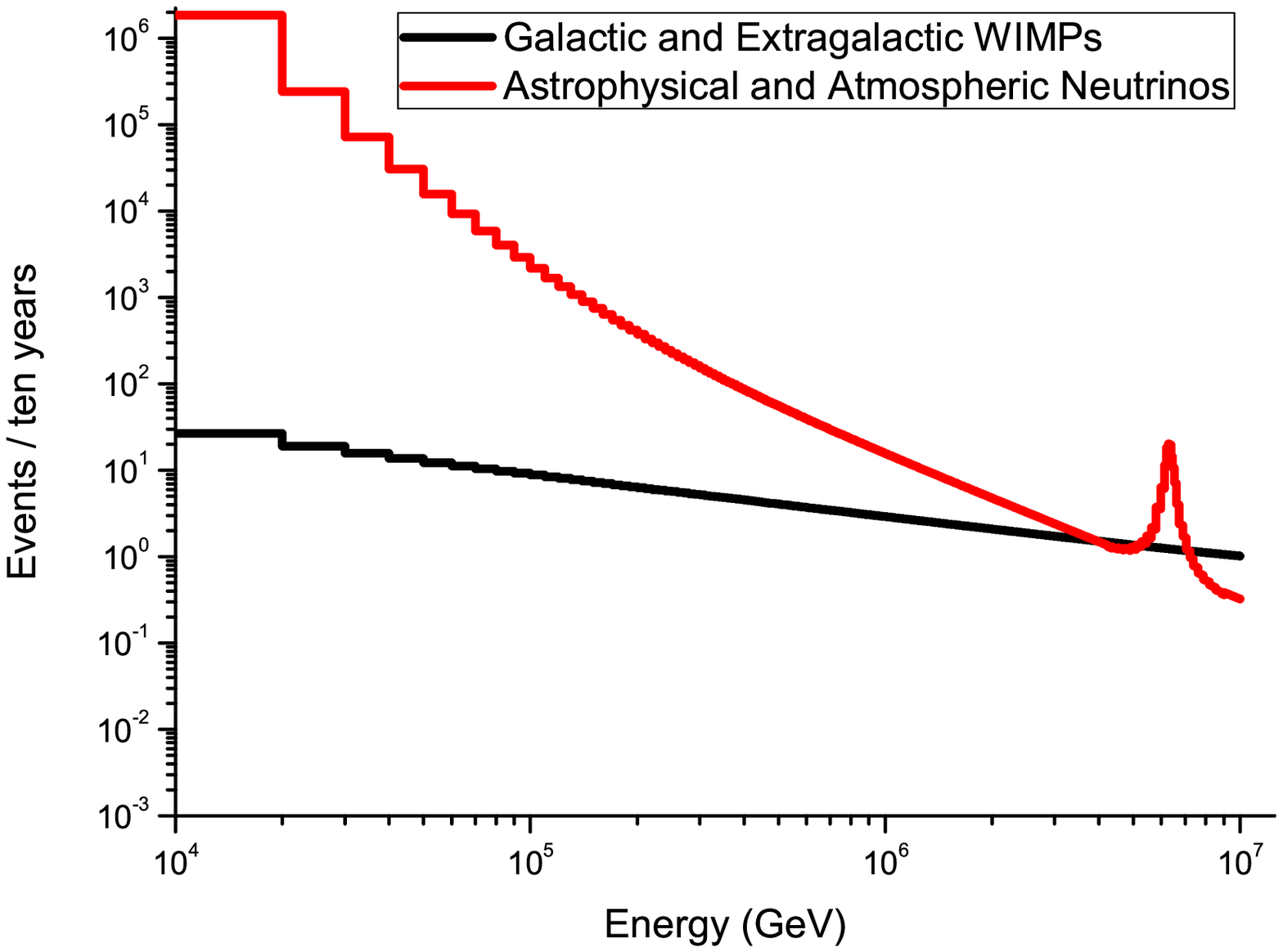}
 \caption{The UHE WIMPs and neutrino event rates are evaluated at $\theta_{max} = \pi$ at IceCube, if $\tau_{\phi}=10^{25}s$ and the lifetime of the IceCube experiment is ten years.}
 \label{fig:180_25_10y}
\end{figure}

\begin{figure}
 \centering
 \includegraphics[width=0.9\textwidth]{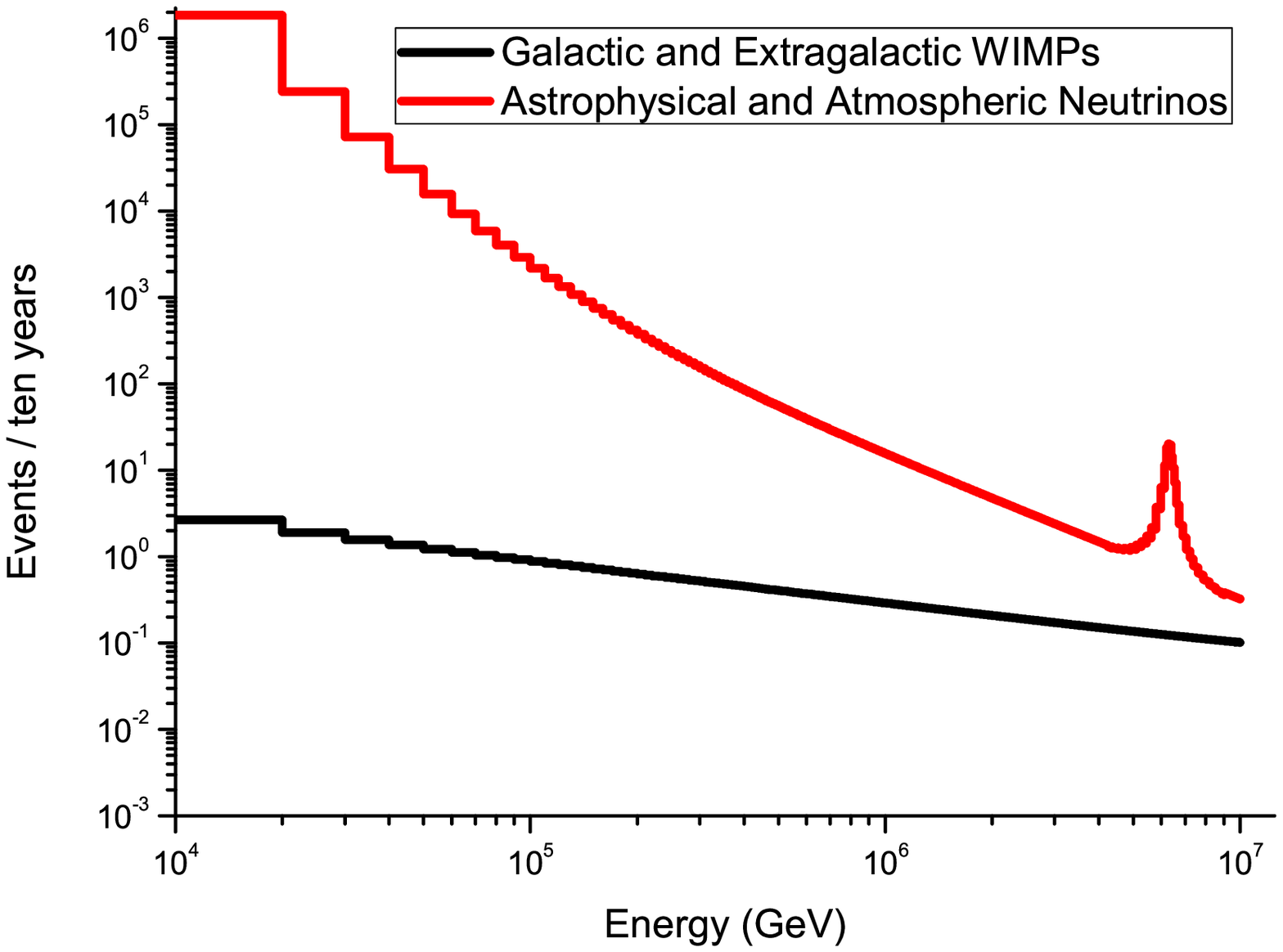}
 \caption{The UHE WIMPs and neutrino event rates are evaluated at $\theta_{max} = \pi$ at IceCube, if $\tau_{\phi}=10^{26}s$ and the lifetime of the IceCube experiment is ten years}
 \label{fig:180_26_10y}
\end{figure}

\begin{figure}
 \centering
 \includegraphics[width=0.9\textwidth]{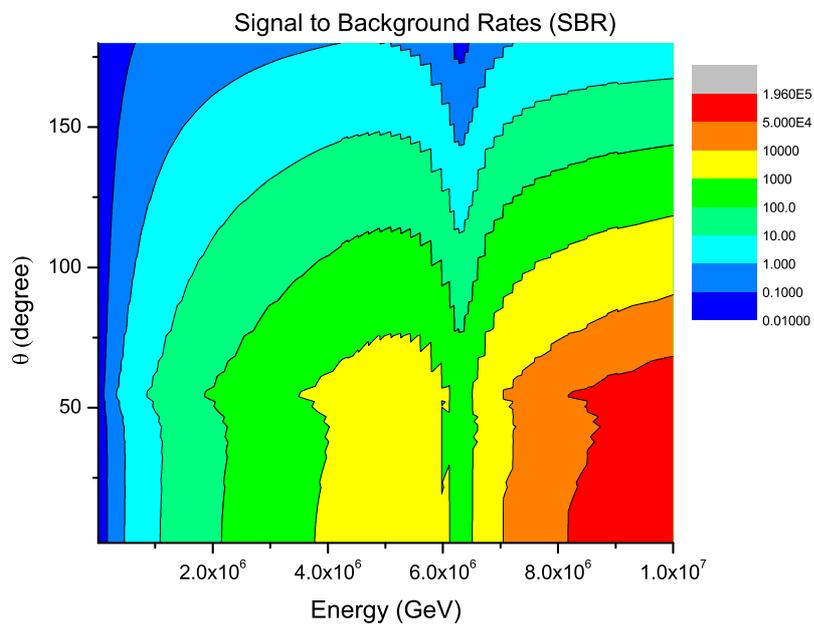}
 \caption{the Signal to Background Rates are evaluated at different energy and different $\theta$, if $\tau_{\phi}=10^{25}s$.}
 \label{fig:sbr_25}
\end{figure}

\end{document}